\documentclass[amsmath,amssymb,floatfix,prb,showpacs,twocolumn]{revtex4-1}
\usepackage{amsmath,amssymb,natbib,bm,graphicx,url,epsfig}
\usepackage[ansinew]{inputenc}
\usepackage{bm}
\usepackage{color}
\usepackage[justification=raggedright,format=plain,indention=.3cm,font=small]{caption}
\usepackage{subcaption}
\usepackage[fleqn,tbtags]{mathtools}

\newcommand{\be}{\begin{equation}}
\newcommand{\ee}{\end{equation}}
\newcommand{\bes}{\begin{equation}\begin{split}}
\newcommand{\ees}{\end{split}\end{equation}}
\newcommand{\bea}{\begin{eqnarray}}
\newcommand{\eea}{\end{eqnarray}}

\newcommand{\bra}[1]{\left\langle \, #1 \,\right|}
\newcommand{\ket}[1]{\left|\, #1 \, \right\rangle}

\DeclareMathOperator{\csch}{csch}

\def\beq{\begin{equation}}
\def\eeq{\end{equation}}
\def\bea{\begin{eqnarray}}
\def\eea{\end{eqnarray}}

\begin{document}

\title{Finite Size Scaling of Topological Entanglement Entropy}
\author{Yuting Wang$^{1}$, Tobias Gulden$^1$, and Alex Kamenev$^{1,2}$}

\affiliation{$^1$School of Physics and Astronomy, University of Minnesota, Minneapolis, MN 55455, USA}
\affiliation{$^2$William I. Fine Theoretical Physics Institute, University of Minnesota, Minneapolis, MN 55455, USA}

\date{\today}
\vspace{0.1cm}

\begin{abstract}
We consider scaling of the entanglement entropy across a topological quantum phase transition in one dimension. The change of the topology manifests itself in a sub-leading term, which scales as $L^{-1/\alpha}$ with the size of the subsystem $L$, here $\alpha$ is the R\'{e}nyi index. This term reveals the universal scaling function $h_\alpha(L/\xi)$, where $\xi$ is the correlation length, which is sensitive to the topological index. 
\end{abstract}

\maketitle

\section{Introduction}\label{Introduction}

Following the pioneering 2006 works of Kitaev and Preskill \cite{KitaevPreskill}, and Levin and Wen \cite{LevinWen}, 
entanglement entropies became a standard and useful tool to study properties of topological systems \cite{Balents,Haldane,GuWen,Pollmann2010,Pollmann2011}. These works found that the entanglement entropy in two-dimensional (2D) systems contains a universal contribution which distinguishes between different topological phases. 
In this case the {\em topological entropy}, given by the logarithm of the quantum dimension, is a contribution at order $L^0$, sub-leading to the generic area law, where $L$ is the subsystem size. It is uniquely related to the long range entanglement and reflects the intrinsic topology of the system. 

In one-dimensional (1D) systems all topological phases are short range entangled, they only differ in their boundary properties \cite{Verstraete,Fidkowski}. When studying the entanglement entropy one introduces virtual cuts, separating a finite-size subsystem from the rest of the system (hereafter considered to be infinite). A topological phase transition changes the properties at these cuts, therefore one expects effects of topology to be detectable through the entanglement entropy. This paper seeks to identify a topological contribution to the entropy in 1D systems and its scaling behavior across the topological quantum phase transition.

Entanglement entropy in 1D systems has been mostly studied in two cases. The first one is a critical system whose continuum limit is described by a conformal field theory (CFT). It was found \cite{Cardy2004,Cardy2009} that the R\'{e}nyi entropies, $S_\alpha$, scale logarithmically with the subsystem size $L$, with a universal coefficient: 
\begin{equation} 
				\label{eq:CFT}
S_\alpha=\frac{c}{6}\left(1+\frac{1}{\alpha} \right)\ln{L},
\end{equation} 
where $c$ is the central charge -- the number of critical degrees of freedom of the system, and $\alpha$ is the R\'{e}nyi index\cite{Renyi}. The second case is an infinite subsystem with a large but finite correlation length $\xi$. Here the leading contribution is logarithmic in correlation length, $S_\alpha=\frac{c}{6}\left(1+\frac{1}{\alpha} \right)\ln{\xi}+\mathrm{const}$. This result can be obtained through transfer matrix methods\cite{Baxterbook,Baxter1993,Peschel1999} or properties of block Toeplitz matrices\cite{Korepin}. To describe the crossover between the two cases Calabrese and Cardy \cite{Cardy2004} connected the two regimes by a universal finite-size scaling function at order $L^0$ which solely depends on the ratio of the subsystem size and the correlation length, $w=L/\xi$, and the R\'{e}nyi index $\alpha$. For $c=1;1/2$ this scaling function was related \cite{Casini2005} to the correlation functions of the sine-Gordon model. The latter in turn may be expressed through solutions of a Painlev\'{e} $V$ equation \cite{Bernard1994}. Importantly, this scaling function does {\em not} contain information about the topological properties of the transition. 

The main question addressed in this paper is if the finite size scaling of the entanglement entropy near a quantum phase transition is sensitive to the change of the topological index in 1D. We show that the answer is affirmative, yet 
qualitatively different from its 2D analog. We find that there is a second scaling function which appears in the 
sub-leading order with the anomalous scaling $\propto L^{-1/\alpha}$ for R\'{e}nyi entropies with $\alpha>1$, and respectively $\ln(L)/L$ for the von Neumann entropy, $\alpha \rightarrow 1$. The overall finite size scaling in the limit 
$L,\xi\to \infty$, while $w=L/\xi$ is fixed, takes the form 
\begin{equation}
 S_\alpha \!=c\left[\frac{1}{6}\! \left(1\!+\!\frac{1}{\alpha} \right)\! \ln \! \left(\frac{L}{a}\right)\!+\! g_{\alpha}(w)\right] + \frac{1}{\alpha\!-\!1}\left(\frac{a}{L}\right)^{1/\alpha} \!h_\alpha(w)
 \label{eq:scaling_conjecture}
\end{equation}
for $\alpha>1$, and 
\begin{equation}
 S_1 = c\left[\frac{1}{3}\ln \left(\frac{L}{a}\right)+ g_1(w)\right] + \frac{a}{L} \ln\left( \frac{L}{a} \right)h_1(w)
 \label{eq:scaling_conjecture2}
\end{equation}
for the von Neumann entropy ($\alpha\to 1$), respectively, where $a$ is a microscopic length scale. The scaling function $g_\alpha(w)$, introduced by Calabrese and Cardy \cite{Cardy2004,Cardy2009} is insensitive to the change of the topological index. It is the next order scaling function $h_\alpha(w)$ which discriminates between phases with different topology. 

Throughout this paper we define topological and non-topological phases by $w=L/\xi$ being positive or negative respectively, with $w=0$ being the critical point. The scaling function $g_\alpha(-w)$ is symmetric in $w$, $g_\alpha(-w)=g_\alpha(w)$, i.e. it does not distinguish between the two phases. On the contrary, as we will show the second scaling function is antisymmetric, $h_\alpha(-w) = -h_\alpha(w)$. Thus it plays the role of a 1D analog of the topological entropy in two dimensions\cite{KitaevPreskill,LevinWen}. 
This topological contribution appears with the anomalous scaling dimension $L^{-1/\alpha}$ (respectively $\ln(L)/L$ for $\alpha=1$). It is worth mentioning that at the critical point the dominant finite-size correction is known \cite{Cardy2010} to be of the order $L^{-2/\alpha}$, while in massive models far away from criticality \cite{CalabresePeschel} the corrections behave as $\xi^{-1/\alpha}$. The scaling function $h_\alpha(w)$ naturally interpolates between these two limits due to its asymptotic behavior $h_\alpha(w)\sim \pm w^{1/\alpha}$ at $|w|\gg 1$, in agreement with the cited behavior in massive models, and $h_\alpha(0)=0$ at criticality. 

Manifestations of the topological nature of 1D transitions in finite size scaling functions were recently studied for some observables. A universal scaling function, distinguishing the trivial and topological phases, was found for the free energy \cite{energy2016}. Other recent studies investigated the fidelity susceptibility and found that it shows sensitivity to the appearing edge states \cite{fidelity2014,fidelity2016}. In these two cases the scaling functions depend on bulk and topological properties of the system and have no apparent symmetry properties. For R\'{e}nyi entanglement entropies the situation appears to be rather different, since there are two independent scaling functions with even and odd parity across the transition. 

The paper is organized as follows: In Section~\ref{Renyi entropies} we review the concepts of entanglement spectrum and R\'{e}nyi entropies, and show how they may be calculated from the correlation matrix. In Section \ref{section3} we briefly review Kitaev model and connections between its entanglement spectrum and scaling functions. Numerical ways to evaluate the R\'{e}nyi entropies and the properties of the two scaling functions are discussed in Section~\ref{sec:scaling functions}. Finally conclusions and open questions are summarized in Section \ref{sec:outlook} . Technical details are relegated to two appendices.

\section{Entanglement spectrum and R\'{e}nyi entropies} 
\label{Renyi entropies} 

We first briefly review the concepts of entanglement spectrum and R\'{e}nyi entropies. The former represents detailed information about the entanglement, while the latter provides a simple measure of entanglement and is commonly used to characterize it. General methods to calculate entanglement spectra and R\'{e}nyi entropies are also introduced below.

Let us assume that the entire system is in a pure state $\ket{\Psi}$, with density matrix $\rho=\ket{\Psi} \bra{\Psi}$. One chooses a part of the system as the subystem $\mathcal{A}$. The information about entanglement between the subsystem $\mathcal{A}$ and the rest of the system, $\mathcal{B}$, is encoded in the reduced density matrix $\rho_\mathcal{A}$. The reduced density matrix is obtained by tracing out all degrees of freedom which are outside of subsystem $\mathcal{A}$, $\rho_\mathcal{A} = \mathrm{Tr}_\mathcal{B} \rho$. One can now introduce the (dimensionless) 
entanglement Hamiltonian $\mathcal{H}_E$ according to, \cite{Peschel01,Peschel03}
\begin{equation}
 \rho_\mathcal{A} = \frac{e^{-\mathcal{H}_E}}{\mathcal{Z}_\mathcal{A}},
 \label{eq:rhoA}
\end{equation}
where $\mathcal{Z}_\mathcal{A}=\mathrm{Tr}(e^{-\mathcal{H}_E})$ is a normalization constant. The eigenvalues of the entanglement Hamiltonian $\mathcal{H}_E$ are commonly referred to as the entanglement spectrum. 

For free fermion models one can write $\mathcal{H}_E=\sum_{i,j}H_{i,j}c_i^\dagger c_j$, where $c_i^\dagger$ is a fermion creation operator on site $i$ and $c_j$ is an annihilation operator on site $j$, and $i,j\in {\cal A}$. One can diagonalize the entanglement Hamiltonian $\mathcal{H}_E$ to get its eigenfunctions $\{\psi_{l}(i)\}$ and corresponding eigenvalues $\{ \epsilon_l \}$. The transformation to new fermion operators $\tilde{c}_l$, $c_i=\sum_l \psi_{l}(i)\tilde{c}_l$ diagonalizes the entanglement Hamiltonian and simultaneously diagonalizes the reduced density matrix:
\begin{equation}
 \rho_{\cal{A}} =\frac{e^{-\sum_{l} \epsilon_l \tilde{c}_{l}^{\dagger} \tilde{c}_{l}}}{\cal{Z}_{\cal{A}}}.
 \label{eq:rhoA2}
\end{equation} 
Using the equation above and $\mathrm{Tr}(\rho_\mathcal{A})=1$, one obtains:
\begin{equation}
 \mathcal{Z}_\mathcal{A}=\prod_l \left(1+e^{-\epsilon_l}\right).
 \label{eq:ZA}
\end{equation}

The entanglement spectrum $\{\epsilon_l\}$ can be obtained from the two-point correlation function of the subsystem $\mathcal{A}$, $C_{i,j}=\langle c_{i}^{\dagger}c_j \rangle$ with $i,j\in\mathcal{A}$. By definition of the reduced density matrix, the two-point correlation function of the subsystem can also be written as $C_{i,j}=\mathrm{Tr}(\rho_{\mathcal A}c_{i}^{\dagger}c_j)$. Using Eq.~\eqref{eq:rhoA2} and Eq.~\eqref{eq:ZA} one gets:
\begin{equation}
 C_{i,j}=\sum_{l} \psi_l^{\ast} (i) \psi_l(j) \frac{1}{e^{\epsilon_l}+1}.
 \label{eq:correlation_function2}
\end{equation}
The correlation matrix $C_{i,j}$ is Hermitian and its eigenvalues are $\lambda_l=\left(e^{\epsilon_l}+1\right)^{-1}$. Inversely, the entanglement spectrum can be calculated from the eigenvalues of the correlation matrix:\cite{Peschel03},
\begin{equation}
 \epsilon_l=\ln \left(\frac{1-\lambda_l}{\lambda_l}\right).
 \label{eq:spectrum_definition}
\end{equation}

R\'{e}nyi entropies $S_{\alpha}$ quantify the amount of quantum entanglement of a subsystem $\mathcal{A}$ with its surroundings $\mathcal{B}$. The R\'{e}nyi entropies between $\mathcal{A}$ and $\mathcal{B}$ are defined through the reduced density matrix:
\begin{equation}
S_{\alpha}=\frac{1}{1-\alpha }\ln \mathrm{Tr} \left(\rho _{\mathcal{A}}\right)^{\alpha },
\end{equation}
where $\alpha$ is the R\'{e}nyi index. The limiting case $\alpha \rightarrow1$ gives the von Neumann entropy $S_1=\mathrm{Tr} \big(\rho _{\mathcal{A}} \ln \rho _{\mathcal{A}}\big)$, which is usually called the entanglement entropy.

By using Eq.~\eqref{eq:rhoA2} and Eq.~\eqref{eq:ZA}, R\'{e}nyi entropies can be written in terms of the entanglement spectrum:
\begin{equation}
 \begin{aligned}
 S_\alpha &= \frac{1}{1-\alpha }\ln\left[\prod_{l} \frac{1+e^{-\alpha \epsilon_l}}{(1+e^{-\epsilon_l})^\alpha}\right]\\
&= \frac{1}{1-\alpha}\sum_{l}\left[\ln\left(1+e^{-\alpha \epsilon_l}\right)-\alpha \ln\left(1+e^{- \epsilon_l}\right)\right].
 \end{aligned}
 \label{eq:entropy}
\end{equation}
Below we use Eq.~\eqref{eq:spectrum_definition} to evaluate the entanglement spectrum for a one dimensional topological model and then apply Eq.~\eqref{eq:entropy} to calculate its R\'{e}nyi entropies.

\section{Entanglement in the Kitaev chain}
\label{section3}

\subsection{The model}

\begin{figure}[h!]
 \includegraphics[width=1.\columnwidth]{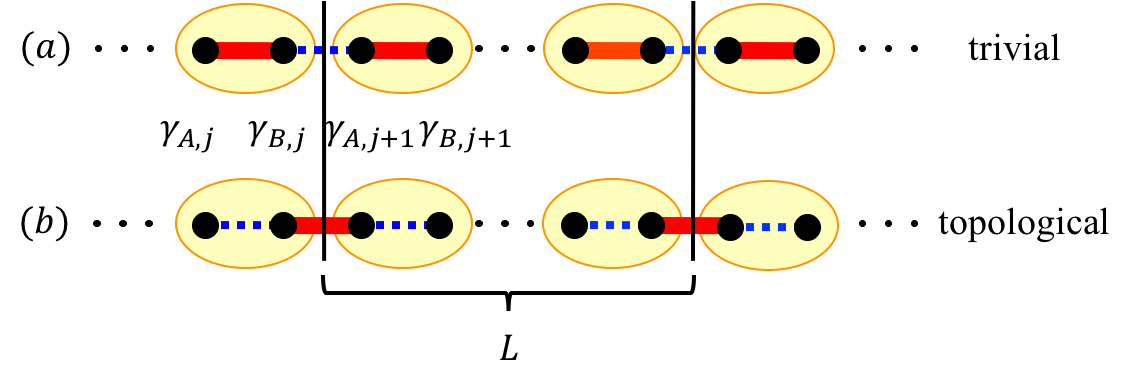}
 \caption{Schematic of the Majorana pairings in (a) non-topological phase $|\mu|>t$ and (b) topological phase $|\mu|<t$ , yellow circles denote that Majorana particles enclosed belong to the same site. Red solid bonds represent strong coupling, blue dashed bonds represent weak coupling. A block of length $L$ is cut out of the infinite system as the subsystem $\cal{A}$, shown by black solid line.}
 \label{fig:KitaevChain}
\end{figure}

Here we employ the Kitaev chain model \cite{Kitaev, Alicea} to study the entanglement spectrum and R\'{e}nyi entropies for a one dimensional topological systems. Its Hamiltonian is 
\begin{equation}
 \mathcal{H}_K = -\mu\sum_{j=1}^N c_j^\dagger c_j - \frac{1}{2}\sum_{j=1}^{N-1}\left(tc_j^\dagger c_{j+1} + \Delta c_jc_{j+1} +h.c. \right),
 \label{eq:Kitaev}
\end{equation}
where $t$ is hopping and $\Delta$ is Cooper pairing amplitudes; $\mu$ is the chemical potential.
The topological properties of the model become apparent when converting the Dirac fermion on each site into a pair of Majorana operators:\cite{Kitaev,Alicea}
\begin{equation}
 c_j = \frac{1}{2}(\gamma_{A,j}+i\gamma_{B,j});\quad c_j^\dagger = \frac{1}{2}(\gamma_{A,j}-i\gamma_{B,j}).
 \label{eq:Majorana_transform}
\end{equation}
The Majorana fermions are their own antiparticles in the sense that $\gamma_{A/B,j}^\dagger=\gamma_{A/B,j}$, and obey the canonical fermionic anti-commutation relations. Fig.~\ref{fig:KitaevChain} shows the Majorana states aligned in a chain. 

When $|\mu|>t$ the bond between Majorana fermions from the same site is stronger than between different sites, resulting in formation of on-site dimers. When $|\mu|<t$ the bond between Majorana fermions from neighboring sites is dominant, which leads to formation of dimers between Majorana fermions $\gamma_{B,j}$ and $\gamma_{A,j+1}$. The Majorana fermions $\gamma_{A,1}$ and $\gamma_{B,N}$ at the ends of the chain remain weakly paired, they form topological zero-energy edge states. The {\em single} fermionic zero energy state, split between the edges of the chain, reflects the degeneracy between even and odd particle number {\em many-body} ground states. 

There are quantum phase transitions between the two phases at $\mu_c=\pm t$. At the critical point the gap closes and the correlation length $\xi \propto (t-|\mu|)^{-1}$ diverges. Away from criticality the correlation length $\xi$ is finite. In the following we identify $\xi > 0$ with the topologically non-trivial state and $\xi < 0$ with the trivial state.

\subsection{Entanglement spectrum}

\begin{figure}[h!]
 \includegraphics[width=1.\columnwidth]{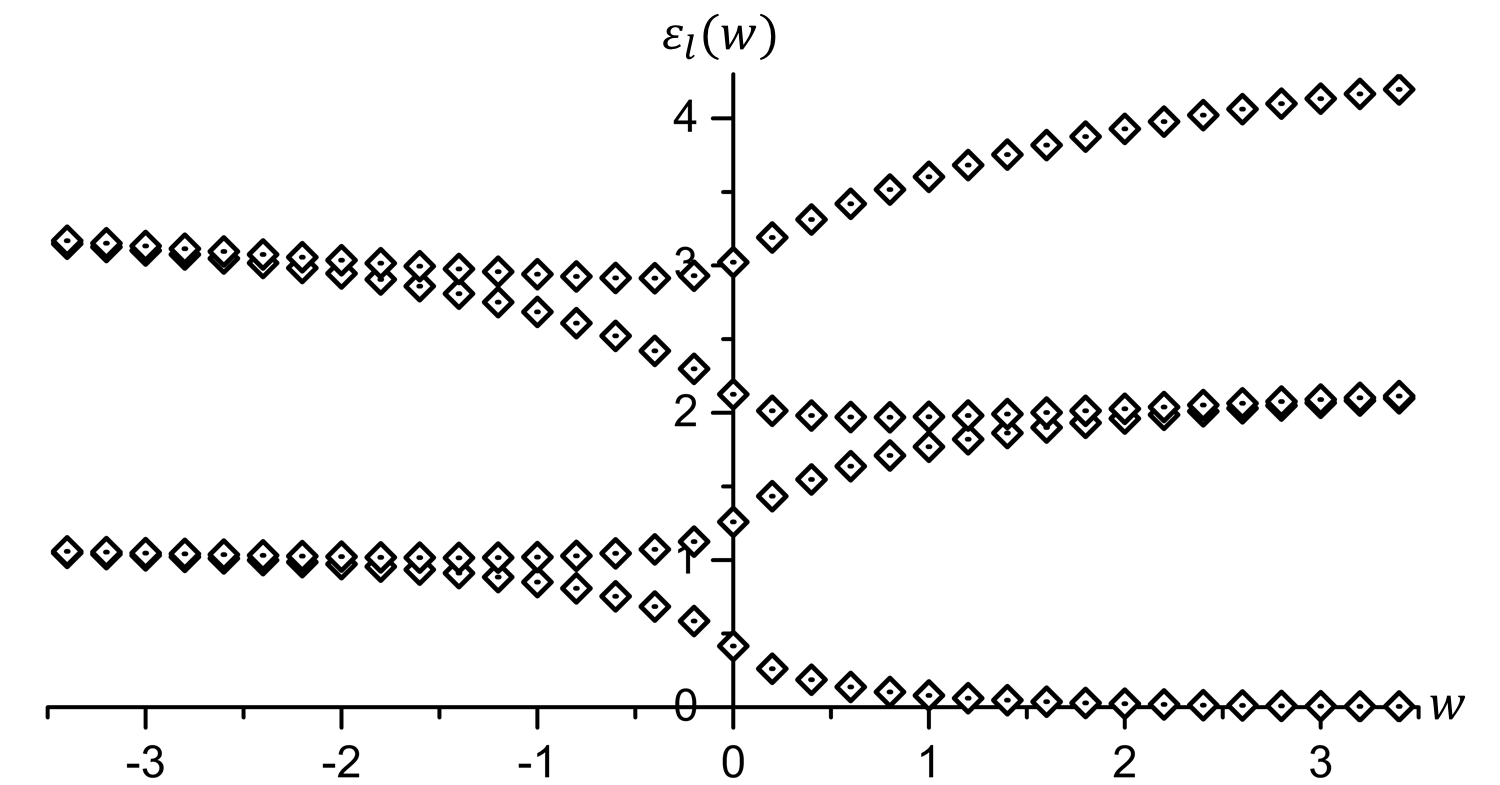}
 \caption{First four levels of the entanglement spectrum as functions of $w=L/\xi$ are shown for the Kitaev model with $\Delta=1$ and $L=5000$. It is apparent that the spectrum is asymmetric in $w$. When going from $w<0$ to $w>0$, the double-degenerate levels first splits into non-degenerate levels and then different neighboring levels pair up again, leaving the lowest level, which approaches zero, unpaired.} 
 \label{fig:EntanglementSpectrum}
\end{figure}

To study entanglement one imagines taking a block of length $L$ in the chain as the subsystem $\cal{A}$, see Fig.~\ref{fig:KitaevChain}. The two-point correlation matrix of the subsystem may be calculated by using the many-body ground state of the model (see Appendix~\ref{app:Correlation functions} for details):
\begin{eqnarray}
  &&C_{2i-1,2j-1}=\bra{gs}{\gamma_{A,i}\gamma_{A,j}}\ket{gs}=\frac{1}{2}\delta_{ij}, \nonumber \\
  &&C_{2i,2j}=\bra{gs}{\gamma_{B,i}\gamma_{B,j}}\ket{gs}=\frac{1}{2}\delta_{ij},      \label{eq:core-matrix}\\
  &&C_{2i-1,2j}=\bra{gs}{\gamma_{A,i}\gamma_{B,j}}\ket{gs}\nonumber \\
 &&\quad =\frac{1}{4\pi} \int_{-\pi}^{\pi}\!\!\! dk\,\, e^{i k (2j+1-2i)}\,\frac{i(t \cos k+\mu)-\Delta \sin k}{\sqrt{(t \cos k+\mu)^2+(\Delta \sin k)^2}}. \nonumber 
\end{eqnarray}
Having the matrix elements of the correlation matrix $C$, one can diagonalize it to find its eigenvalues $\{\lambda_l\}$. Then one can use Eq.~\eqref{eq:spectrum_definition} to calculate the entanglement spectrum $\{\epsilon_l\}$. An example of the entanglement spectrum as a function of the deviation from the criticality is depicted in Fig.~\ref{fig:EntanglementSpectrum}. 

Far from criticality, i.e. $\xi\ll L$, the entanglement spectrum is doubly degenerate. This reflects the fact that the two edges of the subsystem ${\cal A}$ are essentially decoupled and contribute equally to the entanglement spectrum. When going across the phase transition, where $\xi\to\infty$, from the non-topological ($w=L/\xi<0$) to the topological ($w>0$) side, the double-degenerate levels first split and then pair up again with different neighboring levels. On the topological side the lowest level remains non-degenerate and exponentially approaches zero. This zero energy state reflects the even/odd degeneracy of the many-body ground-state of the chain. As a result, the entanglement spectrum is markedly {\em asymmetric} between the two sides. 

The central question of this paper is how the asymmetry of the entanglement spectrum across the topological phase transition is reflected in scaling properties of the corresponding entanglement R\'{e}nyi entropies. Below we demonstrate that the answer to this question is rather subtle and interesting.

To approach this question we first notice that the lowest levels of the entanglement Hamiltonian yield the main contribution to the R\'{e}nyi entropies, cf. Eq.~\eqref{eq:entropy}. One can thus employ a reasonable approximation for these low lying levels to predict large $L$ scaling of the entropies. In the large $|w|$ limit ($L\gg \xi \gg a$), the spectrum is doubly degenerate and equidistant with \cite{Baxterbook,Baxter1993,Peschel1999} 
\begin{equation}
    \epsilon_l(L,w\rightarrow \infty)  = \frac{\pi^2}{\ln\left(\xi/a\right)}
\begin{dcases}
    (l-\delta_{l,\text{even}}),& \!\!\text{ non-topological } \\
    (l-\delta_{l,\,\text{odd}}),& \!\!\text{ topological }
\end{dcases}
\end{equation} 
where $l=1,2,\ldots$ and the Kronecker delta function, $\delta_{l,\text{even/odd}}$ is equal to $1$ if $l$ is even/odd, and $0$ otherwise. On the other hand, at the critical point, the energy levels are non-degenerate and evenly spaced in the large $L$ limit \cite{Peschel04, Peschel09} with spacing $\pi^2/\ln\left(L/a\right)$. Here $a$ is a microscopic length scale which scales as $a\propto1/\Delta$. 
Near criticality, at $|w|\ll 1$, the levels are seen to alternate between descending and ascending ones, see Fig.~\ref{fig:EntanglementSpectrum}. One can thus approximate them as 
\begin{equation}
 \epsilon_l(L,w)= \frac{\pi^2}{\ln(L/a)}\left(l-\frac{1}{2}+(-1)^l \delta_{a}(w)+\delta_{s}(w)\right), 
 \label{eq:spectrum_approx}
\end{equation}
where $\delta_{a}(w)$ is anti-symmetric and alternates between odd and even $l$'s, and $\delta_{s}(w)$ is symmetric and approximately $l$-independent. 

From the model entanglement spectrum (\ref{eq:spectrum_approx}) one can evaluate the $|w|\ll 1$ regime for R\'{e}nyi entropies according to Eq.~(\ref{eq:entropy}). Employing Ramanujan's sum formula\cite{Hardy,MathWorld} (see Appendix~\ref{app:Renyi entropies evaluation} for 
details) we find the leading terms in the limit $L\to \infty$, while $w$ is fixed. The result is given by Eqs.~(\ref{eq:scaling_conjecture}), (\ref{eq:scaling_conjecture2}), where 
\begin{equation}
				\label{eq:g and h}
g_\alpha(w)\propto \delta_{s}(w), \quad\quad h_\alpha(w)\propto \delta_{a}(w);\quad\quad \alpha\geq 1.
\end{equation}
The take-away messages from this exercise is that: (i) the sub-leading term is indeed expected to come with the anomalous scaling dimension $L^{-1/\alpha}$ (for $\alpha>1$); (ii) the scaling functions $g_\alpha(w)$ and $h_\alpha(w)$ import the properties of the underlying entanglement spectrum (at least for small $|w|$) and (iii) it is the sub-leading scaling 
function $h_\alpha(w)$, which discriminates between topological and non-topological phases (the leading scaling function 
$g_\alpha(w)$ appears to be totally symmetric and thus oblivious to the topology). Below we verify and extend these conclusions via extensive numerical simulations.

\section{Scaling functions}   
\label{sec:scaling functions}

\subsection{Numerical analysis}   

R\'{e}nyi entropies $S_\alpha(L,w)$ for the Kitaev model can be calculated from the entanglement spectrum using Eq.~\eqref{eq:entropy}. 
We then perform the scaling analysis by subtracting the critical result, Eq.~(\ref{eq:CFT}), and going to largest available system sizes, to show that $S_\alpha(L,w)-{1\over 12} (1+\alpha^{-1})\ln L=g_\alpha(w)$ is indeed a function of the scaling variable $w$ only. Afterward we go to smaller system sizes to investigate the sub-leading corrections and find that they behave as the last terms in Eqs.~\eqref{eq:scaling_conjecture} and \eqref{eq:scaling_conjecture2}. 

\begin{figure}[h!]
 \includegraphics[width=1.\columnwidth]{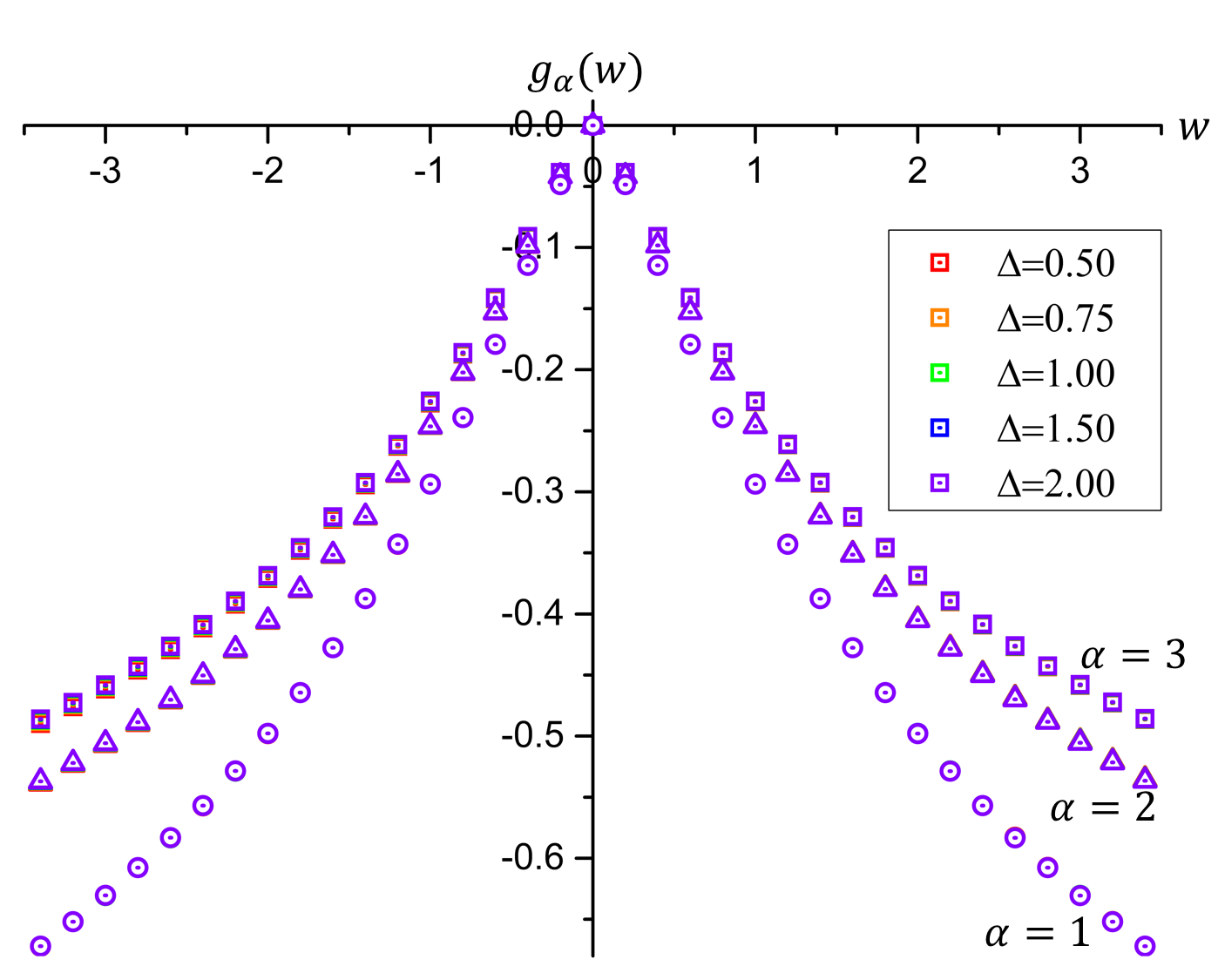}
 \caption{The scaling function $g_\alpha(w)$ is plotted for the Kitaev model for different $\alpha=1, 2, 3$ (circle, triangle, square) and at different $\Delta=0.5, 0.75, 1, 1.5, 2$ (red, orange, green, blue, purple). Here $L_1=5000$, $L_2=4990$ are used. This function is universal for different $\Delta$ up to a constant shift (neglected herein). $g_\alpha(w)$ is symmetric around the topological transition, i.e. it does not depend on the topological phase of the system.}
 \label{fig:g_alpha}
\end{figure}

\begin{figure}[h!]
 \includegraphics[width=1.\columnwidth]{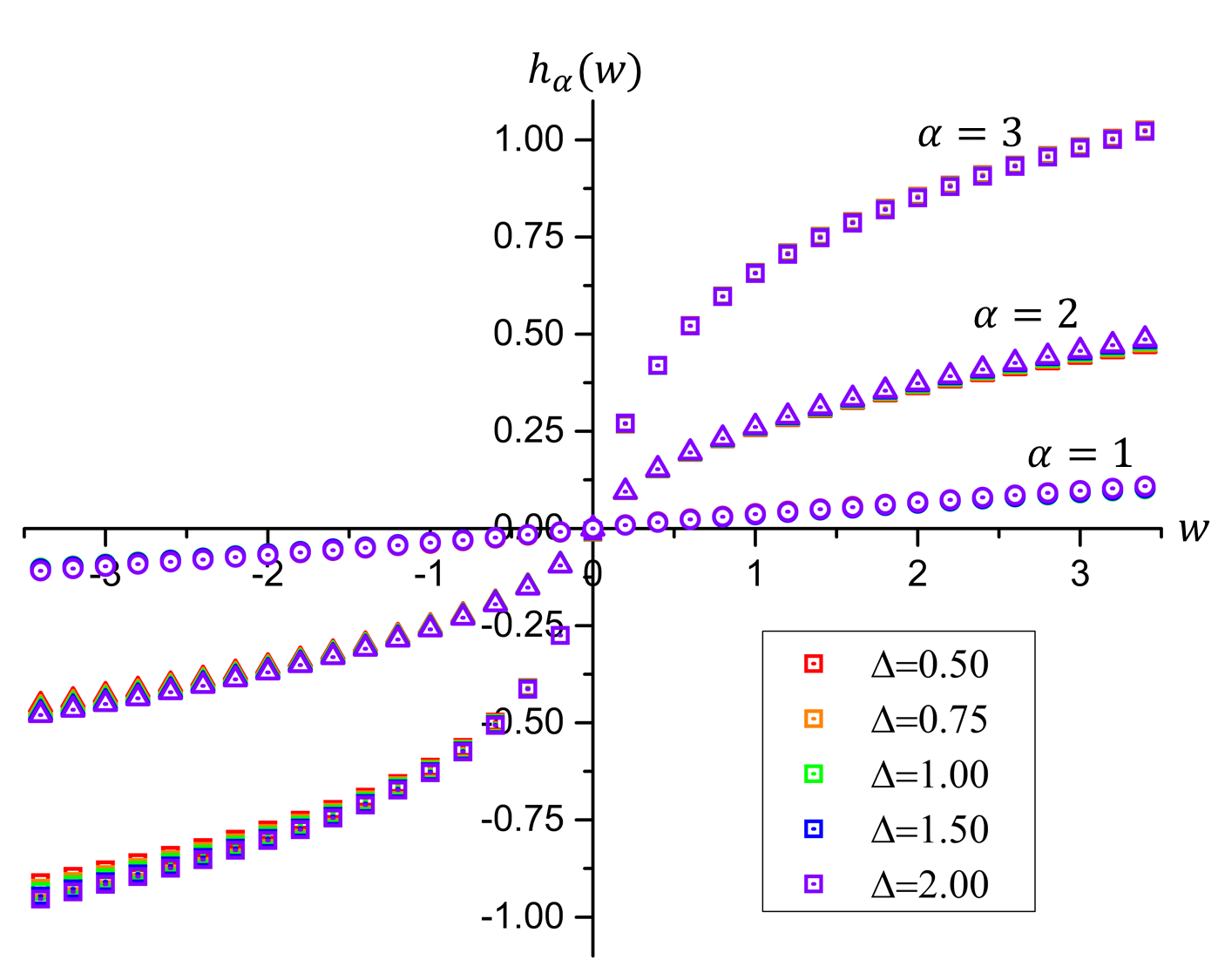}
 \caption{The scaling function $h_\alpha(w)$ is plotted for the Kitaev model for different $\alpha=1, 2, 3$ (circle, triangle, square) and at different $\Delta=0.5, 0.75, 1, 1.5, 2$ (red, orange, green, blue, purple). Here $L_1=5000$, $L_2=4990$ are used. $h_\alpha(w)$ is anti-symmetric around the topological transition, i.e. it depends on the topological phase of the system.}
 \label{fig:h_alpha}
\end{figure}

Once the scaling form, Eqs.~\eqref{eq:scaling_conjecture}, \eqref{eq:scaling_conjecture2}, is established, the higher quality data are obtained in the following way: we eliminate function $g_{\alpha}(w)$ by subtracting $S_{\alpha}$ at subsystem size $L_2$ from that at a different subsystem size $L_1$, keeping $w$ fixed:
\begin{multline}
S_\alpha(L_1,w)-S_\alpha(L_2,w)=\frac{c}{6}\left(1+\frac{1}{\alpha} \right) \ln \left(\frac{L_1}{L_2}\right)\\+\frac{1}{1-\alpha}\left[\left(\frac{a}{L_1}\right)^{1/\alpha}\!\!-\left(\frac{a}{L_2}\right)^{1/\alpha}\right]h_\alpha(w).
\end{multline}
After re-organizing the above equation, one gets:
\begin{equation}
 h_\alpha(w)=(1-\alpha)\frac{S_\alpha(L_1,w)\!-\!S_\alpha(L_2,w)\!-\!\frac{c}{6}\left(1\!+\!\frac{1}{\alpha} \right) \ln\! \left(\frac{L_1}{L_2}\right)}{\left(\frac{a}{L_1}\right)^{1/\alpha}\!\!-\left(\frac{a}{L_2}\right)^{1/\alpha}}
 \label{eq:h_alpha}
\end{equation}
To get $g_{\alpha}(w)$, one subtracts the leading logarithmic term and the topological scaling function, Eq.~\eqref{eq:h_alpha}, from R\'{e}nyi entropies at subsystem size $L_1$:
\begin{multline}
 g_\alpha(w)=\frac{1}{c}\left [ S_\alpha(L_1,w)-\frac{c}{6}\left(1+\frac{1}{\alpha} \right) \ln \left(\frac{L_1}{a}\right)\right. \\ -\frac{S_\alpha(L_1,w)-S_\alpha(L_2,w)-\frac{c}{6}\left(1+\frac{1}{\alpha} \right) \ln \left(\frac{L_1}{L_2}\right)}{1-\left(\frac{L_1}{L_2}\right)^{1/\alpha}} \bigg].
 \label{eq:g_alpha}
\end{multline}
Using Eq.~\eqref{eq:h_alpha} and Eq.~\eqref{eq:g_alpha} for the various subsystem sizes $L_1$ we show that the data indeed converge to the universal (i.e. $\Delta$ independent) scaling functions. These scaling functions $g_\alpha(w)$ and $h_\alpha(w)$ for $\alpha=1, 2, 3$ are shown in Fig.~\ref{fig:g_alpha} and Fig.~\ref{fig:h_alpha} correspondingly. 


\subsection{Scaling function $g_\alpha(w)$}

The most remarkable fact about entanglement entropy scaling in 1D is that the $L^0$ scaling function $g_\alpha(w)$ is entirely symmetric between topological and non-topological sides of the transition. This is surprising at the first glance, since the entanglement spectrum $\{\epsilon_l\}$ is markedly asymmetric as seen in Fig.~\ref{fig:EntanglementSpectrum}. Yet, once the spectrum is used to calculate the entropy according to Eq.~(\ref{eq:entropy}), the result is fully symmetric to the $\ln L$ and $L^0$ order for {\em any} $\alpha$.  
This is also what follows from the calculations based on the model spectrum (\ref{eq:spectrum_approx}), as seen from 
Eq.~(\ref{eq:g and h}). Therefore to this order the entanglement entropies are completely insensitive to the change of the topology between the two sides of the transition. This should be contrasted with the 2D case where the constant $L^0$ term carries a hallmark of the topological nature of the phase\cite{KitaevPreskill,LevinWen,Balents,Haldane,GuWen,Pollmann2010,Pollmann2011}. Apparently the situation in 1D is qualitatively different and one should look for other signatures of the topological transition. 

Though, oblivious to the topology, the scaling function $g_\alpha(w)$ is still a fascinating object and we review some of its properties here for completeness. It was proposed \cite{Cardy2004,Casini2005} to be universal for different models up to a non-universal constant shift. Our calculations support this conclusion, since there is no difference between different values of $\Delta$ in the Kitaev model. We have also investigated the Su-Schriffer-Heeger model\cite{sshPRL,sshTopol} and found that $g_\alpha(w)$ is the same as in the Kitaev model. Nevertheless, the function $g_\alpha(w)$ still depends on index $\alpha$. As mentioned above, it is fully symmetric $g_\alpha(w)=g_\alpha(-w)$ within the accuracy of our simulations. 

\begin{figure}[h!]
 \includegraphics[width=.9\columnwidth]{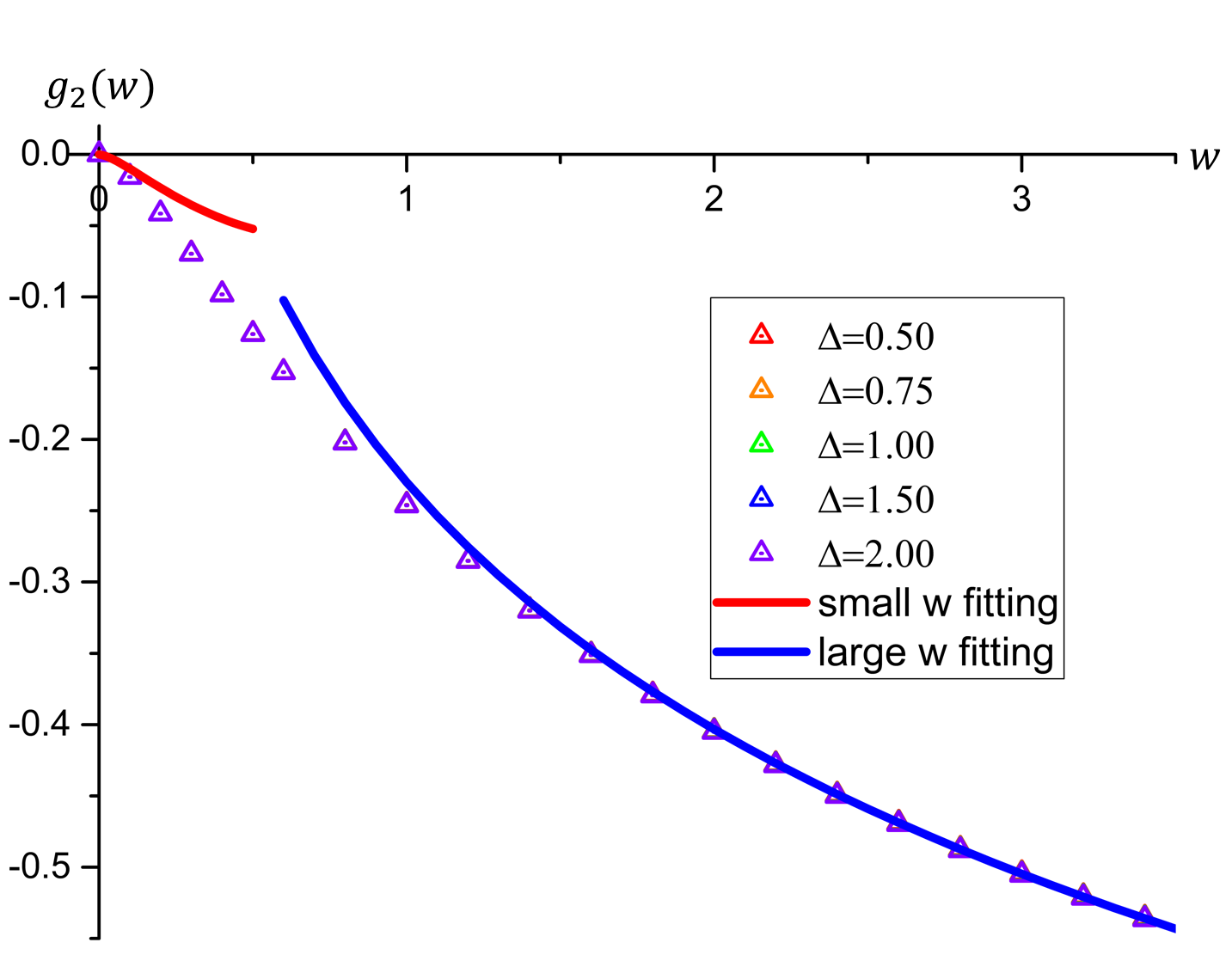}
 \caption{The scaling function $g_\alpha(w)$ is plotted for the Kitaev model at different $\Delta=0.5, 0.75, 1, 1.5, 2$ (red, orange, green, blue, purple). Here $\alpha=2$, and $L_1=5000$, $L_2=4990$ are used. $g_\alpha(w)$ is symmetric around the topological transition so we only plot the function in the topological phase. Red solid line shows the small $w$ approximation, Eq.~\eqref{eq:g_small_fitting}. Blue line shows the large $w$ approximation, Eq.~\eqref{eq:g_large_fitting}.}
 \label{fig:g_alpha2_entire_fitting}
\end{figure}

For large $w$, i.e. $L\gg \xi$, the entanglement entropies must approach an $L$-independent limit, which indicates:
\begin{equation}
 g_\alpha(w)=-\frac{1}{6}(1+\frac{1}{\alpha})\ln|w|; \quad \quad  |w|\gg 1. 
 \label{eq:g_large_fitting}
\end{equation}
This is indeed what the numerics show, see Fig.~\ref{fig:g_alpha2_entire_fitting}. 
For small $|w|\ll 1$ Ref.~[\onlinecite{Casini2005}] gives an approximation of $g_\alpha(w)$ as  
\begin{equation}
 g_\alpha(w)=-\frac{1}{6}\left(1+\frac{1}{\alpha}\right)\left(\frac{1}{2}w^2\ln^2|w|-\frac{1}{2}w^2\ln |w|+\frac{w^2}{4}\right).
 \label{eq:g_small_fitting}
\end{equation}
The comparison of this asymptotic result with the numerical data is also shown in Fig.~\ref{fig:g_alpha2_entire_fitting} for $\alpha=2$. 

\subsection{Topological entanglement Entropy $h_\alpha(w)$}

\begin{figure}[h!]
 \includegraphics[width=1.\columnwidth]{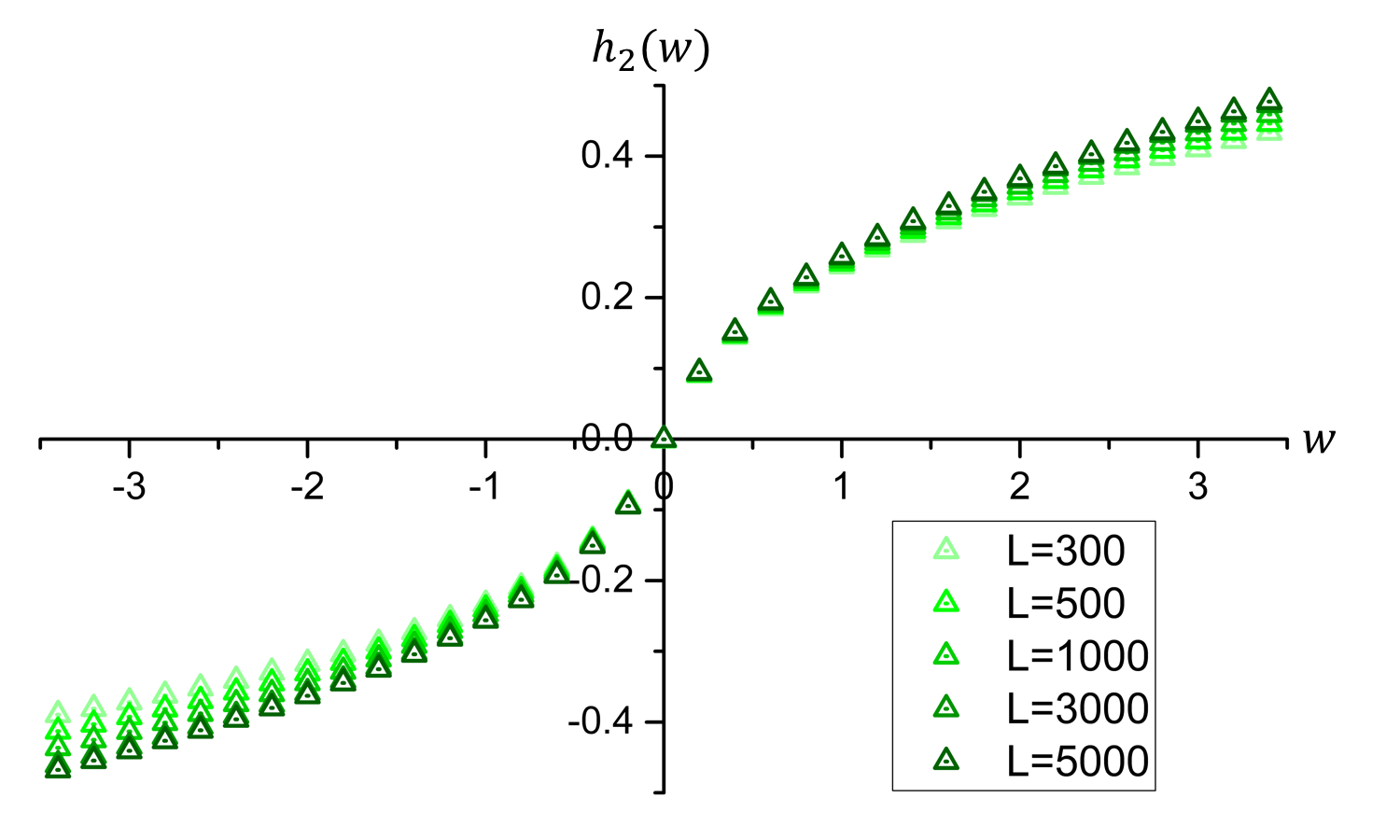}
 \caption{The function $h_\alpha(w)$ is calculated from size $L_1=300$ and $L_2=290$, $L_1=500$ and $L_2=490$, $L_1=1000$ and $L_2=990$, $L_1=3000$ and $L_2=2990$, $L_1=5000$ and $L_2=4990$ (light to dark green) for the Kitaev model. Here $\alpha=2, \Delta=1$. The function $h_\alpha(w)$ is convergent when the subsystem size increases, thus it is a scaling function.}
 \label{fig:h_alpha_scaling}
\end{figure}

\begin{figure}[h!]
 \includegraphics[width=.9\columnwidth]{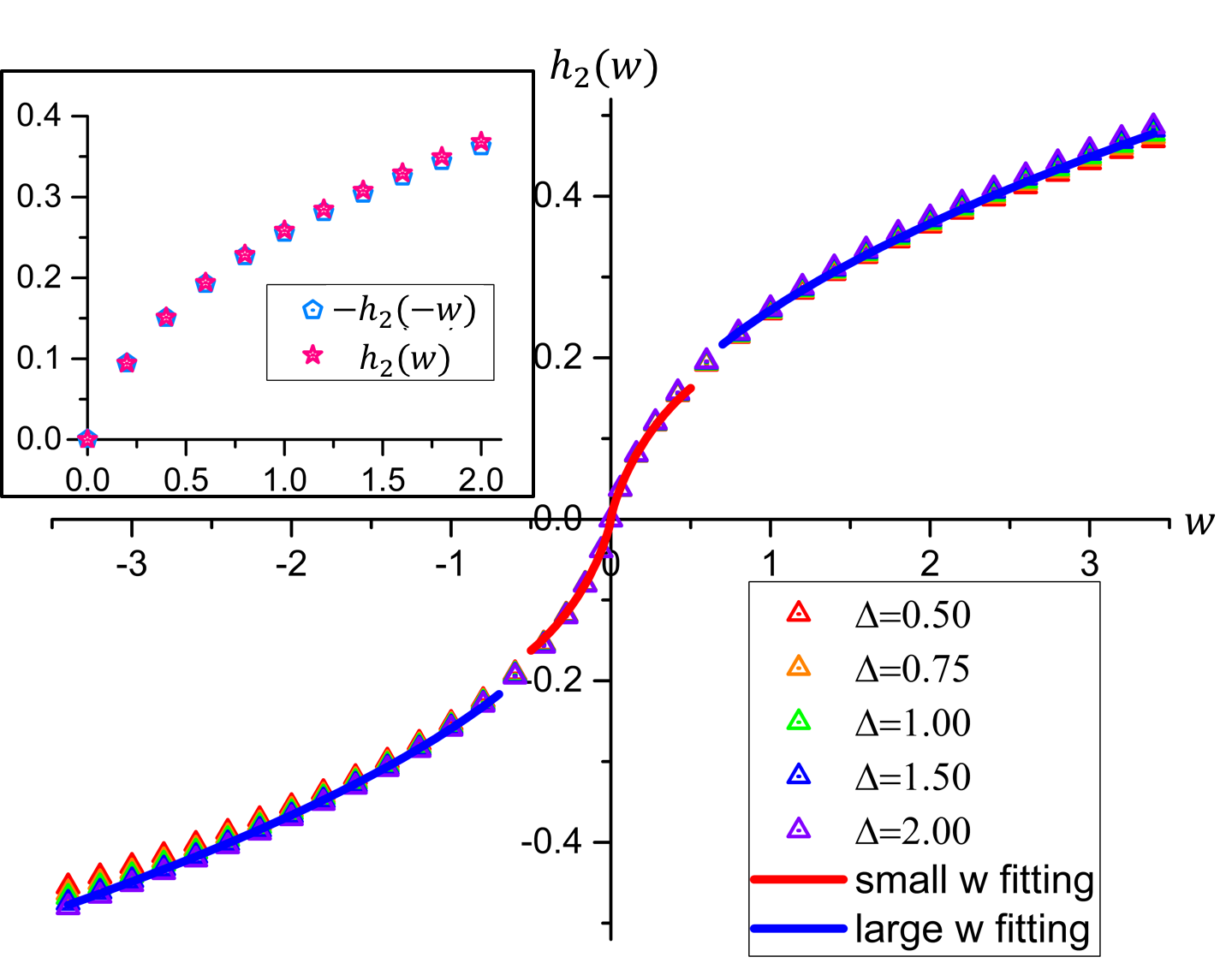}
 \caption{The scaling function $h_\alpha(w)$ is universal for Kitaev model at different $\Delta=0.5, 0.75, 1, 1.5, 2$ (red, orange, green, blue, purple). Here $\alpha=2$ and $L_1=5000$, $L_2=4990$ are used. The function is antisymmetric (inset in the panel) around the phase transition and depends on the topological state of the system. Red solid line is the fitting of small $w$ approximation evaluated by Ramanujan's sum formula, see Appendix~\ref{app:Renyi entropies evaluation}. Blue line is the fitting of large $w$ approximation which is proportional to $w^{1/\alpha}$ with $\alpha=2$.}
 \label{fig:h_alpha2}
\end{figure}

The main result of this paper is that in 1D the topological information is encoded in the sub-leading term $\propto L^{-1/\alpha}$. For $\alpha >1$ it comes with the new scaling function $h_\alpha(w)$, which discriminates between the topological and the non-topological sides of the transition. Our data (see Fig.~\ref{fig:h_alpha_scaling} and Fig.~\ref{fig:h_alpha2}) show the following key features of this function: (i) $h_{\alpha}(w)$ is indeed a scaling function -- this can be seen from the data collapse as the system is increased at fixed $w$; (ii) although the prefactor depends on a non-universal microscopic length scale $a$, the scaling function $h_{\alpha}(w)$ itself is universal, i.e. independent of the model parameters, such as $\Delta$; (iii) in agreement with Eq.~(\ref{eq:g and h}), $h_{\alpha}(w)$ is an anti-symmetric function of its argument. 

The latter observation implies $h(0)=0$, i.e. corrections of order $L^{-1/\alpha}$ are {\em absent} at the conformal point $w=0$. This is consistent with Refs.~[\onlinecite{Cardy2010,CalabreseEssler}], who found that the leading finite size correction to the conformal result scales as $ L^{-2/\alpha}$ and thus $ L^{-1/\alpha}$ must be non-existent at $w=0$. 


For $L\gg \xi$ one expects the entropy to be $L$-independent. This immediately implies that $h_\alpha(w) \sim w^{1/\alpha}$ for $|w|\gg1$. Then, together with the prefactor $L^{-1/{\alpha}}$ , the sub-leading correction to the R\'enyi entropies is proportional to $\xi^{-1/\alpha}$. This agrees with the correction to R\'{e}nyi entropies in the region far away from criticality obtained by Calabrese and Peschel \cite{CalabresePeschel}. It also agrees with our numerical data, as shown in Fig.~\ref{fig:h_alpha2}. 

For small $w$ one can use the model (\ref{eq:spectrum_approx}), which predicts that the scaling functions mirror the small $w$ behavior of the low-lying entanglement levels, Eq.~(\ref{eq:g and h}). Since the correlation matrix elements exhibit $w\ln |w|$ non-analyticity, which may be deduced from Eq.~(\ref{eq:core-matrix}), this non-analytic behavior shows up in $\epsilon_l(w)$ functions. Indeed, by fitting the lowest entanglement levels, see Fig.~\ref{fig:EntanglementSpectrumSmall} in Appendix \ref{app:Renyi entropies evaluation}, we find:
\begin{equation}
 \begin{aligned}
 &\delta_{a}(w) \approx -0.32w \ln |w|+0.45w, \\
 &\delta_{s}(w) \approx 0.024w^2\ln^2|w|-0.069w^2 \ln |w|+0.056w^2. 
 \end{aligned}
 \label{eq:spectrum_fitting}
\end{equation}
This suggests non-analytic behavior of the scaling function $h_\alpha(w)\propto w\ln|w|$. This is indeed consistent with the data, see Fig.~\ref{fig:h_alpha2}.

\section{Discussion and Outlook}
\label{sec:outlook}

We have shown that the entanglement entropy of 1D symmetry-protected topological models carries the information about 
change of the topological index across the quantum phase transition. Contrary to 2D systems, such information resides in the sub-leading correction with the anomalous $\propto L^{-1/\alpha}$ scaling dimension, here $\alpha\geq 1$ is the R\'enyi index and $L$ the subsystem size. This correction comes with the new scaling function $h_\alpha(w)$, where
$w=L/\xi$ and the double scaling limit: $L\to \infty; \xi\to\infty$, while $w=$const, is assumed. We found that $h_\alpha(w)$ is a universal function (up to a multiplicative factor) and uncovered its asymptotic behavior in the limit of large and small argument. It is this scaling function which discriminates between topological and non-topological sides of the quantum phase transition. 

These observations pose a number of open questions. One of them is an analytic evaluation of the anomalous scaling function $h_\alpha(w)$. We notice that the $L^0$ scaling function $g_\alpha(w)$, through a mapping onto a continuum bosonized theory, is connected to a known correlation function of the bosonic sine-Gordon model \cite{Casini2005}. It is a fascinating question whether a similar construction is capable of revealing $h_\alpha(w)$. One reason to be cautious about this approach is that the term in question must be proportional to $a^{1/\alpha}$, where $a$ is a microscopic length scale not present explicitly in a continuum theory. Another open question is universality of both scaling functions beyond $c=1$ and $c=1/2$ models. In that case, other numerical methods may be used to tackle the problem.\cite{Pollmann2009} These questions may become subjects of future works.

\section{Acknowledgements} The authors gratefully acknowledge useful discussions with Fiona Burnell and Marc Schulz.
This work was supported by NSF grant DMR-1608238.

\appendix
\section{Correlation functions}
\label{app:Correlation functions}

In this Appendix we derive the two-point correlation functions for the Kitaev model. Employing Eq.~\eqref{eq:Majorana_transform} one transforms the Hamiltonian, Eq.~\eqref{eq:Kitaev}, into the Majorana basis:
\begin{multline}
 \mathcal{H}_K = \frac{i}{4}\sum_j^N\left[-2\mu \gamma_{A,j}\gamma_{B,j}+(t-\Delta)\gamma_{B,j}\gamma_{A,j+1}\right.\\\left.+(t+\Delta)\gamma_{A,j}\gamma_{B,j+1}\right].
\end{multline}
One can then diagonalize the above Hamiltonian to obtain its eigenvalues $E^\pm(k)$ and corresponding eigenfunctions $\Psi_{\sigma,k}^\pm(j)$:
\begin{equation}
\begin{dcases}
   E^\pm(k)\!\!\!\!&=\pm \sqrt{(t\cos k+\mu)^2+(\Delta\sin k)^2}\\
   \Psi_{\sigma,k}^\pm(j)\!\!\!\!&=\sigma {1\over {\sqrt{2} N}}\,e^{ikj} e^{-i\sigma\phi_k}
   \label{eq:wave_function}
\end{dcases},
\end{equation}
where $\pm$ refers to upper and lower bands, $\sigma=\pm1$ are used to label the Majorana fermions A/B on each lattice site and $e^{-2i\phi_k}=\left[i(t \cos k+\mu)-\Delta \sin k\right]/\sqrt{(t \cos k+\mu)^2+(\Delta \sin k)^2}$. 

In the many body ground state, all states in the lower band are occupied:
\begin{equation}
 \ket{gs}=\prod_{k\in gs} \tilde{\gamma}_{A,k} \tilde{\gamma}_{B,k} \ket{0},
\end{equation}
where $\ket{0}$ is the vacuum. With 
\begin{equation}
 \gamma_{\sigma,j}=\sum_k \Psi_{\sigma,k}^-(j)\tilde{\gamma}_{\sigma,k},
\end{equation}
the corresponding ground state correlation functions take the form: 
\begin{equation}
 \bra{gs}\gamma_{\sigma,i}^\dagger\gamma_{\sigma',j}\ket{gs}=\sum_{k\in gs}\Psi_{\sigma,k}^{-*}(i)\Psi_{\sigma',k}^-(j).
\end{equation}
Finally, inserting Eq.~\eqref{eq:wave_function} into the above equation and taking the continuum limit, one obtains the two-point correlation functions shown in Eqs.~(\ref{eq:core-matrix}).

\section{Evaluating R\'enyi entropies}
\label{app:Renyi entropies evaluation}

\begin{figure}[h!]
 \includegraphics[width=1.\columnwidth]{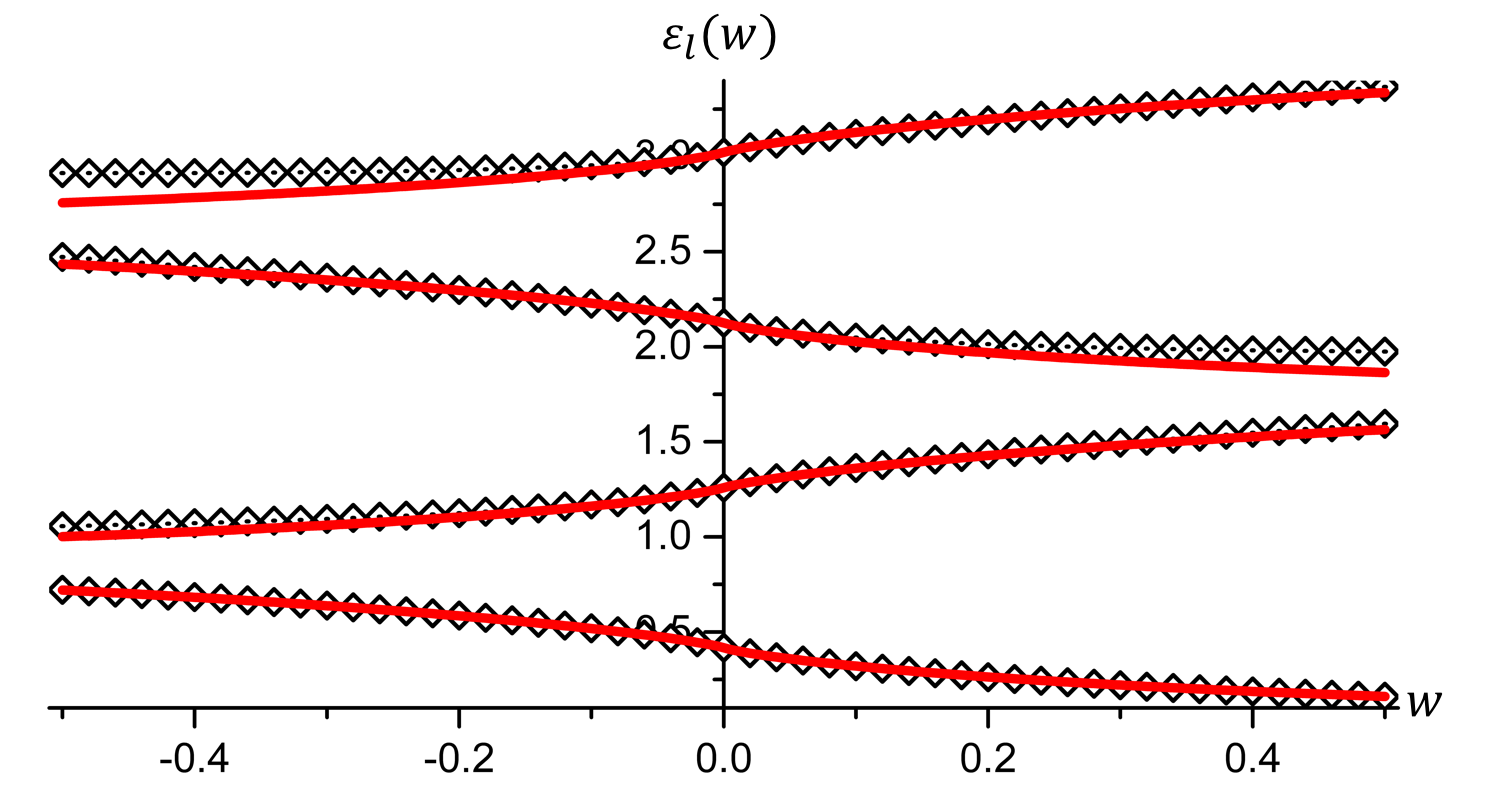}
 \caption{First four levels of entanglement spectrum as function of $w=L/\xi$ are shown for the Kitaev model at $\Delta=1$ and $L=5000$. Red solid lines are small $w$ fitting of the entanglement spectrum using Eq.~\eqref{eq:spectrum_approx} and Eq.~\eqref{eq:spectrum_fitting}.}
 \label{fig:EntanglementSpectrumSmall}
\end{figure}

In this Appendix we evaluate the R\'{e}nyi entropies for $|w|\ll1$ by applying Eq.~\eqref{eq:entropy} to the approximated entanglement spectrum Eq.~\eqref{eq:spectrum_approx}. Equations \eqref{eq:spectrum_approx} and \eqref{eq:spectrum_fitting} are used to fit the lowest levels in the entanglement spectrum. The fitting is shown in Fig.~\ref{fig:EntanglementSpectrumSmall}.

In order to carry out the summation of alternating spectrum in Eq.~\eqref{eq:entropy}, the first thing to do is to separate the spectrum Eq.~\eqref{eq:spectrum_approx} into odd and even levels and relabel them with $n=(l-1)/2$ for odd $l$ and $n=(l-2)/2$ for even $l$:
\begin{equation}
\begin{aligned}
 &\epsilon_n^o = 2\epsilon n+\epsilon \left[\frac{1}{2}+\delta_s(w)-\delta_a(w)\right],\\
 &\epsilon_n^e = 2\epsilon n+\epsilon \left[\frac{3}{2}+\delta_s(w)+\delta_a(w)\right], 
 \label{eq:relabelled spectrum}
\end{aligned}
\end{equation}
where $n=0,1,2,\ldots$, and $\epsilon=\pi^2/ \ln(L/a)$.

The two sets of odd and even levels in the entanglement spectrum are semi-infinite in $n$ and evenly spaced for all small $w$, so we can apply Ramanujan's sum formula\cite{Hardy,MathWorld} which is similar to the familiar Poisson summation:
\begin{equation}
 \sqrt{\gamma}\left[\sum_{n=0}^\infty\phi(n\gamma)-\frac{1}{2}\phi(0)\right] = \sqrt{\beta}\left[\sum_{n=0}^\infty\psi(n\beta)-\frac{1}{2}\psi(0)\right],
 \label{eq:Ramanujan}
\end{equation}
where $\beta\gamma=2\pi$ and
\begin{equation}
 \psi(x) = \sqrt{\frac{2}{\pi}}\int_0^\infty\phi(t)\cos(xt)dt.
 \label{eq:Fourier}
\end{equation}

For Eq.~\eqref{eq:entropy}, defining the functions:
\begin{equation}
 \phi_\alpha^{o/e}(t) = \ln\left(1+p_\alpha^{o/e}e^{-t}\right),
  \label{eq:phi}
\end{equation}
with 
\begin{equation}
\begin{aligned}
 p_\alpha^{o}&=e^{-\alpha\epsilon\left[1/2+\delta_s(w)-\delta_a(w)\right]},\\
 p_\alpha^{e}&=e^{-\alpha\epsilon\left[3/2+\delta_s(w)+\delta_a(w)\right]},
\end{aligned}
\end{equation}
we can rewrite the summation, Eq.~\eqref{eq:entropy} as:
\begin{multline}
 S_\alpha = \frac{1}{1-\alpha}\sum_{n=0}^\infty \left[\phi_\alpha^o(2\alpha\epsilon n) - \alpha\phi_1^o(2\epsilon n)\right. \\\left. + \phi_\alpha^e(2\alpha\epsilon n) - \alpha\phi_1^e(2\epsilon n) \right].
 \label{eq:Renyi_phi}
\end{multline}
The Fourier transform of the functions $\phi_\alpha^{o/e}(t)$ gives:
\begin{equation}
 \psi_\alpha^{o/e}(x) 
         = -\sqrt{\frac{2}{\pi}}\sum_{m=1}^\infty\frac{\left(-p_\alpha^{o/e}\right)^m}{m^2+x^2}.
 \label{eq:psi}
\end{equation}
One can apply the sum formula Eq.~\eqref{eq:Ramanujan} to Eq.~\eqref{eq:Renyi_phi} to get
\begin{multline}
\!\!\!\!\!\!S_\alpha\!=\! \frac{1}{1-\alpha}\! \left\{ \!\sqrt{\frac{\pi}{2}} \frac{1}{\alpha \epsilon}\!\left[\sum_{n=0}^\infty \psi_\alpha^{o,e}\!\!\left(\frac{\pi n}{\alpha\epsilon}\right)\!\!-\!\frac{1}{2}\psi_\alpha^{o,e}(0)\right]\!\!+\!\frac{1}{2}\phi_\alpha^{o,e}(0)\right. \\ \left.-\sqrt{\frac{\pi}{2}} \frac{\alpha}{\epsilon} \! \left[\sum_{n=0}^\infty \psi_1^{o,e}\!\!\left(\frac{\pi n}{\epsilon}\right)\!-\!\frac{1}{2}\psi_1^{o,e}(0)\right]\!\!-\!\frac{\alpha}{2}\phi_1^{o,e}(0)\!\right\}.
\end{multline}
By using the summation over $n$:
\begin{equation}
\begin{aligned}
 &\sum_{n=0}^\infty\psi_\alpha^{o/e}\left(\frac{\pi n}{\alpha\epsilon}\right)=-\sqrt{\frac{2}{\pi}}\sum_{m=1}^\infty \left(-p_\alpha^{o/e}\right)^m \sum_{n=0}^\infty \frac{1}{m^2+\left(\frac{\pi n}{\alpha\epsilon}\right)^2}\\
 &=\frac{1}{2}\psi_\alpha^{o,e}(0)-\sqrt{\frac{2}{\pi}}\sum_{m=1}^\infty\frac{\left(-p_\alpha^{o/e}\right)^m}{2m}\alpha\epsilon\coth(m\alpha\epsilon),
\end{aligned}
\end{equation}
we arrive at:
\begin{multline}
\!\!\!\!\!\!S_\alpha\!=\!\frac{1}{1-\alpha}\!\!\sum_{m=1}^\infty \!\!\frac{(-1)^m}{m} \!\!\!\left\{\!\alpha \csch(m\epsilon)e^{\!-m\epsilon\delta_s}\!\cosh\!\!\left[\!m\epsilon\!\left(\!\frac{1}{2}+\delta_a\!\right)\!\right]
\right.\\ \left.-\!\csch (m\alpha\epsilon)e^{\!-m\alpha\epsilon\delta_s}\!\cosh\!\!\left[\!m\alpha\epsilon\!\!\left(\!\frac{1}{2}+\delta_a\!\right)\!\!\right]\!\right\}.
 \label{eq:Renyi_sum}
\end{multline}

To find the leading contributions for R\'enyi entropies, we expand the above result up to power $\epsilon^0$:
\begin{eqnarray}
 S_\alpha &=& \frac{\pi^2}{12\epsilon}\left(1+\frac{1}{\alpha}\right)-\delta_s(w)\ln2 \nonumber\\
          &=& \frac{1}{12}\left(1+\frac{1}{\alpha}\right)\ln\left(\frac{L}{a}\right)-\delta_s(w)\ln2.
 \label{eq:Renyi_leading}
\end{eqnarray}
The leading logarithmic term is the well-known critical result for a system with conformal charge $c=\frac{1}{2}$.\cite{Cardy2004} The second term, which is the main correction to the critical result, depends only on the symmetric contribution in the entanglement spectrum $\delta_s(w)$. 

To find the sub-leading contribution to R\'{e}nyi entropy we note in Eq.~\eqref{eq:spectrum_approx} that for small $|w|\ll1$ the antisymmetric perturbation dominates, $\delta_a(w)\gg\delta_s(w)$. Neglecting the symmetric part $\delta_s(w)$ in Eq.~\eqref{eq:relabelled spectrum} for now, and defining the function:
\begin{equation}
 \phi_\alpha(t) = \ln\left(1+e^{-\alpha t}\right),
\end{equation}
one can rewrite R\'enyi entropies, Eq.~\eqref{eq:entropy} as:
\begin{multline}
 S_\alpha = \frac{1}{2(1-\alpha)}\sum_{n=-\infty}^\infty \left[\phi_\alpha(2n\epsilon+\eta^o) - \alpha\phi_1(2n\epsilon+\eta^o)\right. \\\left. + \phi_\alpha(2n\epsilon+\eta^e) - \alpha\phi_1(2n\epsilon+\eta^e) \right],
\end{multline}
where $\eta^o=\epsilon(1/2-\delta_a(w))$ and $\eta^e=\epsilon(3/2+\delta_a(w))$.
Then instead of Ramanujan's sum, we apply the generalized Poisson summation to the above equation:
\begin{equation}
\sqrt{\gamma}\sum_{n=-\infty}^\infty\phi(n\gamma+\eta)= \sqrt{\beta}\sum_{n=-\infty}^\infty\psi(n\beta)e^{i n \beta\eta},
 \label{eq:Poisson}
\end{equation}
then the following result is obtained:
\begin{multline}
\!\!\!\!\!\!S_\alpha = \frac{\pi^2}{12\epsilon}\left(1+\frac{1}{\alpha}\right)\\+\frac{1}{1-\alpha}\sum_{n=1}^\infty\! \cos\!\left[\!\pi n\!\!\left(\frac{1}{2}-\delta_a(w)\!\right)\!\right]\!\frac{\alpha\csch\!\left(\!\frac{n\pi^2}{\epsilon}\!\right)\!\!-\!\csch\!\left(\!\frac{n\pi^2}{\alpha\epsilon}\!\right)}{n}.
\end{multline}
The first term is the critical result and identical to Eq.~\eqref{eq:Renyi_leading}. To find the main dependence on $L$ in the second term we use
\[
 \csch\left(\frac{n\pi^2}{\alpha\epsilon}\right)=\csch\left(\frac{n}{\alpha}\ln\frac{L}{a}\right)\approx2\left(\frac{L}{a}\right)^{-n/\alpha}.
\]
The $n=1$ term gives the main sub-leading contribution to the R\'{e}nyi entropy. For $\delta_a(w)\ll1$, the sub-leading correction is proportional to $\delta_a(w)L^{-1/\alpha}$ for $\alpha>1$ and $\delta_a(w)\ln (L)/L$ for the particular case $\alpha \rightarrow 1$.


\end{document}